\documentclass{llncs}

\usepackage{graphicx} 
\usepackage{pdflscape} 
\usepackage{booktabs} 
\usepackage[left=2.5cm,right=3cm,top=3cm,bottom=2.5cm]{geometry}
\usepackage{fancyhdr} 
\usepackage[utf8]{inputenc}
\usepackage{float}
\usepackage{datetime}
\usepackage{natbib}
\usepackage{setspace}
\usepackage{amsmath}
\usepackage{hyperref}
\usepackage{verbatim}
\usepackage{algorithm}
\usepackage{algpseudocode}
\usepackage{marginnote}
\usepackage{rotating}

\usepackage{amsfonts}

\algnewcommand\algorithmicinput{\textbf{Input:}}
\algnewcommand\Input{\item[\algorithmicinput]}
\algnewcommand\algorithmicoutput{\textbf{Output:}}
\algnewcommand\Output{\item[\algorithmicoutput]}
\usepackage{comment}

\usepackage{bbm}
\usepackage{pifont}
\usepackage{subfig}
\usepackage{xr}

\usepackage{xcolor}

\hypersetup{
    colorlinks,
    linkcolor={red!50!black},
    citecolor={red!50!black},
    urlcolor={blue!80!black}
}

\begin{document}

\title{\textit{pMSE} Mechanism: Differentially Private Synthetic Data with Maximal Distributional Similarity}
\titlerunning{\textit{pMSE} Mechanism}  
%
\author{Joshua Snoke \and Aleksandra Slavkovi\'{c}}
\authorrunning{Snoke and Slavkovi\'{c}} 
%
%
\institute{Department of Statistics \\Pennsylvania State University, University Park, PA 16802, USA\\
\email{\{snoke,sesa\}@psu.edu}}

\maketitle            

\begin{abstract}
We propose a method for the release of differentially private synthetic datasets. In many contexts, data contain sensitive values which cannot be released in their original form in order to protect individuals' privacy. Synthetic data is a protection method that releases alternative values in place of the original ones, and differential privacy (DP) is a formal guarantee for quantifying the privacy loss. We propose a method that maximizes the distributional similarity of the synthetic data relative to the original data using a measure known as the \textit{pMSE}, while guaranteeing $\epsilon$-differential privacy. Additionally, we relax common DP assumptions concerning the distribution and boundedness of the original data. We prove theoretical results for the privacy guarantee and provide simulations for the empirical failure rate of the theoretical results under typical computational limitations. We also give simulations for the accuracy of linear regression coefficients generated from the synthetic data compared with the accuracy of non-differentially private synthetic data and other differentially private methods. Additionally, our theoretical results extend a prior result for the sensitivity of the Gini Index to include continuous predictors.
\end{abstract}
\section{Introduction}
In many contexts, researchers wish to gain access to data which are restricted due to privacy concerns. 
While there are many proposed methods for allowing researchers to fit models or receive query responses from the data, there are other cases where either due to methodological familiarity or modeling flexibility, researchers desire to have an entire dataset rather than a set of specific queries. This paper proposes a method for releasing synthetic datasets under the framework of $\epsilon$\emph{-differential privacy}, which formally quantifies and guarantees the privacy loss from these releases. 

Differential privacy(DP), originally proposed by \cite{dwork2006calibrating},
is a formal method of quantifying the privacy loss related to any release of information based on private data; for a more in-depth review see \cite{dwork2014algorithmic}, and for a non-technical primer see \cite{nissim2017differential}. Since its inception, DP has spawned a large literature in computer science and some in statistics. It has been explored in numerous contexts such as machine learning algorithms (e.g., \cite{blum2005practical,kasiviswanathan2011can,kifer2012private}), categorical data (e.g., \cite{barak2007privacy,li2010optimizing}), dimension reduction (e.g., \cite{chaudhuri2012near,kapralov2013differentially}), performing statistical analysis (e.g., \cite{wasserman2010statistical,karwa2016inference}), and streaming data (e.g., \cite{dwork2010pan}), to name only a few applications.


While DP is a rigorous risk measure, it has lacked flexible methods for modeling and generating synthetic data. Non-differentially private synthetic data methods (e.g., see \cite{RRR_2003,reiter_cart,Reiter_2010,drechsler2011synthetic,raab2017practical}) while not offering provable privacy, provide good tools for approximating accurate generative models reflecting the original data. Our proposed method maintains the flexible modeling approach of synthetic data methodology, and in addition maximizes a metric of distributional similarity, the \textit{pMSE}, between the released synthetic data and the original data, subject to satisfying $\epsilon$-DP.
We also do \textit{not} require one of the most common DP assumptions concerning the input data, namely that it is bounded, and we do not limit ourselves to only categorical or discrete data. 
We find that our method produces good results in simulations, and it provides a new avenue for releasing DP datasets for potentially a wide-range of applications.

Our specific contributions are: 
(1) the combination of the flexible synthetic data modeling framework with the guarantee of $\epsilon$-DP, (2) the relaxation of DP assumptions concerning boundedness or discreteness of the input data, (3) the embedding of a metric within our mechanism guaranteeing maximal distributional similarity between the synthetic and original data, and (4) a proof for the sensitivity bound of the Gini Index for CART models in the presence of continuous predictors.

The rest of the paper is organized as follows. Section 2 gives a review of important results from differential privacy that we rely on and provides a review of related methods to ours which we use for comparison in our simulation study. Section 3 details our proposed methodology for sampling differentially private data with maximal distributional similarity. Section 4 provides theoretical results for the privacy guarantees of our algorithm. Section 5 shows simulations that support our theoretical findings and provide an empirical estimate of the privacy loss under standard computational practices. Section 6 provides simulation results for the comparison of the accuracy of linear regression coefficients calculated from our method and other DP methods. Section 7 details conclusions and future considerations.

\section{Differential Privacy Preliminaries}
\label{sec:backgroundDP}
Differentially Privacy is a formal framework for quantifying the disclosure risk associated with the release of statistics or raw data derived from private input data. The general concept relies on defining a randomized algorithm which has similar output probabilities regardless of the presence or absence of any given record, as formalized in Definition \ref{def:eDP}. We replicate the definitions and theorems here using notation assuming $X \in \mathbb{R}^{n \times q}$ and $\theta \in \mathbb{R}^{k}$. $X$ is an original data matrix, and we wish to release instead a private version, $X^s$, with the same dimension. $\theta$ is a vector of parameters corresponding to a chosen parameteric model, which can be used to generate synthetic data that reflects the generative distribution of $X$. Further restrictions may be placed on $\theta$ depending on the parametric model.
\begin{definition}[\cite{dwork2006calibrating}]\label{def:eDP}
A randomized algorithm, $\mathcal{M}$, is $\epsilon$-Differentially Private if for all $\mathcal{S} \subseteq Range(\mathcal{M})$ and for all $X, X^\prime$ such that $\delta(X, X^\prime) = 1$:
\[
\frac{P(\mathcal{M}(X^\prime) \in \mathcal{S})}{P(\mathcal{M}(X) \in \mathcal{S})} \leq exp(\epsilon).
\]
\end{definition}
The privacy is controlled by the $\epsilon$ parameter, with values closer to zero offering stronger privacy. Relaxations of $\epsilon$-DP have been proposed to reframe the privacy definition or to improve the statistical utility. A few examples are approximate differential privacy (also known as $(\epsilon,\delta)$-DP, see \cite{dwork2006our}), probabilistic differential privacy (\cite{machanavajjhala2008privacy}), on-average K-L privacy (\cite{wang2016average}), or concentrated privacy (\cite{dwork2016concentrated,bun2016concentrated}). We do not cover these relaxtions further, since our work relies on the stronger $\epsilon$-DP.

A general example of an $\epsilon$-DP mechanism, which produces private outputs, is the Exponential Mechanism defined by \cite{mcsherry2007mechanism}; see Definition \ref{def:expMech}. For a given $\theta$ that we wish to release, this mechanism defines a distribution from which private samples, $\tilde\theta_i$, can be made and released in place of the original vector.
\begin{definition}[\cite{mcsherry2007mechanism}]\label{def:expMech}
The mechanism that releases values with probability proportional to
\[
exp\Bigg(\frac{-\epsilon \; u(X, \theta)}{2 \; \Delta u}\Bigg),
\]
where $u(X, \theta)$ is a quality function that assigns values to each possible output, $\theta$, satisfies $\epsilon$-DP. 
\end{definition}
$\Delta u$ is the global sensitivity, and it is defined as the greatest possible change in the $u$ function for any two inputs, differing in one row. Note that some definitions of DP use the addition or deletion of a row, but here we assume $X$ and $X^\prime$ have the same dimension. More formally:
\begin{definition}[\cite{dwork2006calibrating}]\label{def:sens}
For all $X, X^\prime$ such that $\delta(X, X^\prime) = 1$, the Global Sensitivity (GS) of a function $u:\mathbb{R}^{n \times q} \to \mathbb{R}_{\geq 0}$ is defined as:
\[
\Delta u = \underset{\theta}{sup} \underset{\delta(X, X^\prime) = 1}{sup} |u(X, \theta) - u(X^\prime, \theta)|
\]
\end{definition}
We also rely on two theorems, known as post-processing and sequential composition. The first, stated in Proposition \ref{prop:post-proc}, says that any function applied to the output from a differentially private algorithm is also differentially private. We use this to generate synthetic data based on private parameters, rather than directly generating differentially private data.
\begin{proposition}[\cite{dwork2006calibrating, nissim2007smooth}]\label{prop:post-proc}
Let $\mathcal{M}$ be any randomized algorithm, such that $\mathcal{M}(X)$ satisfies $\epsilon$-differential privacy, and let $g$ be any function. $g(\mathcal{M}(X))$ also satisfies $\epsilon$-differential privacy.
\end{proposition}
Sequential composition, stated in Theorem \ref{thm:comp}, says that for multiple outputs from a differentially private algorithm, one must compose the $\epsilon$ values for each output to produce the overall privacy loss of the process. We need to compose the privacy if we make multiple draws of private parameters from which we produce multiple private synthetic datasets. We may want to release multiple synthetic datasets for better accuracy in estimates based on the data.
Estimates based on multiple datasets are calculated using appropriate combining rules; see \cite{raab2017practical} for details. 
\begin{theorem}[\cite{mcsherry2009privacy}]\label{thm:comp}
Suppose a randomized algorithm, $\mathcal{M}_j$, satisfies $\epsilon_j$-differential privacy for $j = 1,...,q$. The sequence $\mathcal{M}_j(X)$ carried out on the same $X$ provides $(\Sigma_j \; \epsilon_j)$-differential privacy.
\end{theorem}

These theorems and definitions lay the groundwork for our method. Next, we give a brief overview of some related methods to ours which we use for comparison in our simulations in Section \ref{sec:simLM}.

\subsection{Review of Related Methods}
\label{sec:relatedMethods}
A number of different methods have been proposed for releasing differentially private synthetic datasets, although only a few are focused on real-valued, $n \times q$ matrices. \cite{bowen2016comparative} proposed drawing data from a noisy Bayesian Posterior Predictive Distribution (BPPD), and \cite{wasserman2010statistical} generate data from smooth histograms. \cite{bowen2016comparative} provides a fairly comprehensive list of DP synthetic data methods. Many of these methods are limited to specific data types, such as categorical data (e.g., \cite{abowd2008protective,charest2011can}), or network data (e.g., \cite{karwa2016inference,karwa2017sharing}), or they are computational infeasibile for data with a reasonable number of dimensions, such as the Empirical CDF (\cite{wasserman2010statistical}), NoiseFirst/StructureFirst (\cite{xu2013differentially}), or the PrivBayes (\cite{zhang2017privbayes}), though some recent work has proposed ways to reduce the computation time and improve the accuracy (\cite{li2018highdimHist}).

The noisy BPPD method from \cite{bowen2016comparative} is similar to ours in the sense that focuses on drawing generative model parameters from a noisy distribution and then using these private parameters generates private synthetic data according to post-processing. In this case private parameters are drawn from posterior distribution $f(\theta|s^*)$ where $s$ is the Bayesian sufficient statistic and $s^*$ is the statistic perturbed according to the Laplace mechanism (e.g., see \cite{dwork2006calibrating}).
\cite{bowen2016comparative} recommend drawing multiple sets of private parameters and producing a synthetic dataset for each one, which requires composing $\epsilon$.

The smooth histogram method from \cite{wasserman2010statistical} works non-parametrically by binning the data, using these bins to estimate a consistent density function, and applying smoothing to the function which guarantees DP before drawing new samples. The DP smooth histrogram is defined as:
\begin{align}
\hat f^*_K(\textbf{x}) = (1 - \lambda)\hat f_K(\textbf{x}) + \lambda\Omega
\end{align}
where $K$ is the total number of bins, $\Omega = \Bigg(\prod_{j=1}^p (c_{j1} - c{j0}) \Bigg)^{-1}$, $\lambda \geq \frac{K}{K + n(e^{\epsilon/n} - 1)}$, and $\hat f_K(\textbf{x})$ is a mean-squared consistent density histogram estimator. This method does not need to generate multiple datasets, since it is not redrawing model parameters, and accordingly does not need to split $\epsilon$ across multiple synthetic datasets.

However, both of these methods require bounded data. This can be seen explicitly in the smooth histogram formulation where we assume the $j_{th}$ variable has bounds $[c_{j0}, c_{j1}]$. We also need to assume bounds in order to create (and sample from) a finite $K$ bins. The boundedness assumption is less explicit in the noisy BPPD method, but it comes into play when calculating the sensitivity of the statistics. 
If the data were unbounded, the sensitivity could be infinite, which would mean we have to sample the noise from a Laplace distribution with infinite variance.

We want to avoid this assumption in our method because assuming bounds is problematic when it comes to approximating the underlying generative distribution. Continuous data may be naturally unbounded, and at best in many real data scenarios we do not know what the bounds should be. 
If our bounds are too loose, we introduce more noise than necessary through the privacy process, limiting our accuracy. On the other hand, we introduce bias if we set the bounds too low because that truncates the generative distribution below its true range. We further explore the effect of these assumed bounds using a simulation study in Section \ref{sec:simLM}. 

Our method avoids the bounding problem by sampling from a distribution that shrinks in probability as we move to the tails. We do not bound the sample space, but we have low probability of sampling values which are far from the truth. This allows us to produce private data that accurately reflects the natural range of the data. We describe this further in Section \ref{sec:pMSE}.

Furthermore, the smooth histogram method suffers from computational limitations as the number of variables increases, since it divides the data matrix into bins, the number of which grows $O(p^p)$. 
Our method has computational limitations too, though of a different nature, which we discuss further in Section \ref{sec:synExpMech}.
One nice aspect of the noisy BPPD method is that it is computationally fast.

Finally, our method improves over these methods by incorporating a measure of distributional similarity on the resulting synthetic data. The noisy BPPD and smooth histogram add noise to the generative model for the synthetic data. These mechanisms concern only minimizing the noise added to the parameters, but they guarantee nothing concerning the data generated using these parameters. Our method on the other hand, finds the private parameters which can be used to generate synthetic data that will maximize the distributional similarity with respect to the original data. Regardless of the dataset, our method finds the best private parameters for an assumed synthesizing model. Sections \ref{sec:synExpMech} and \ref{sec:sensBound} give a detailed explaination and theoretical results for our method.

\section{Sampling Differentially Private Synthetic Data via the \textit{pMSE} Mechanism}
\label{sec:synExpMech}
We propose to release a DP synthetic data matrix, $X^s$, in place of the original data $X$ and with the same dimension, $n \times q$. In practice it is infeasible to sample a matrix of this size using the Exponential Mechanism. 
More simply, we sample private model parameters and then generate synthetic data based on these noisy parameters. We know from the results on post-processing, see Proposition \ref{prop:post-proc}, that data generated based on these DP parameters are also DP. 
Based on the exponential mechanism we draw DP samples using
\begin{align}\label{eq:expMech}
f(\theta) \propto exp\Bigg(\frac{-\epsilon \; u(X, \theta)}{2\Delta u}\Bigg), 
\end{align}
where $\epsilon$ is our privacy parameter, $u(X, \theta)$ is our quality (or utility) function, and $\Delta u$ is the sensitivity of the quality function. In practice we use a Monte Carlo Markov Chain (MCMC) approach, using the Metropolis algorithm to generate samples from this unnormalized density, since we do not know the value of the $u$ function a priori. Next we define our quality function and derive a bound on its sensitivity. 

\subsection{Defining the Quality Function using the \emph{pMSE}}
\label{sec:pMSE}
We base our quality function on the \emph{pMSE} statistic developed in \cite{WooProp} and \cite{snoke2018utility}:
\[
\emph{pMSE}=\frac{1}{N}\Sigma^N_{i=1}({\hat{p}_i - 0.5})^2,
\]
where $\hat p_i$ are predicted probabilities (i.e., propensity scores) from a chosen classification algorithm.  Algorithm \ref{algDPpmse} gives the steps for calculating the \emph{pMSE} statistic. The \textit{pMSE} is simply the mean-squared error of the predicted probabilities from this classification task, and it is a metric to assess how well we are able to discern between datasets based on a classifier. If we are unable to discern, the two datasets have high distributional similarity. A $pMSE = 0$ means every $\hat p_i = 0.5$, and it implies the highest utility. There has been much work dedicated to tuning models for out-of-sample prediction, but for our purposes we only use the classifier to get estimates of the in-sample predicted probabilities. 
\begin{algorithm}
\caption{General Method for Calculating the \emph{pMSE}}
\label{algDPpmse}
  \begin{algorithmic}[1]
    \item stack the $n$ rows of original data $X$ and the $n$ rows of masked data $X^s$ to create $X^{comb}$ with $N=2n$ rows
    \item add an indicator variable, $I$, to $X^{comb}$ s.t. $I = \{1: x^{comb}_i \in X^s\}$
    \item fit a model to predict $I$ using predictors $Z = f(X^{comb})$.
    \item find predicted probabilities of class $1$, $\hat p_i$, for each row of $X^{comb}$ 
    \item obtain the \emph{pMSE} $=\frac{1}{N}\Sigma^N_{i=1}({\hat{p}_i - 0.5})^2$ 
  \end{algorithmic}
\end{algorithm} 

To make our quality function a function of $\theta$, the vector of parameters we wish to sample, we use the expected value of the \emph{pMSE} given $\theta$, i.e.,
\begin{align}
u(X, \theta) = E[pMSE(X, X^s_\theta) | X, \theta]
\end{align}
where $X^s \sim g(\theta)$. In practice we approximate this by generating $m$ datasets for a given set of parameters and calculate the average \emph{pMSE} across each data set. This approximation does not affect the privacy guarantee (as shown in the proof for Theorem \ref{theorem:sensBound}), but for accuracy $m$ should be large enough to give satisfactory results for estimating the expected value of $u(X, \theta)$. 

As we mentioned before, this quality function makes no assumptions concerning whether the original data are categorical, discrete, or continuous. Secondly, because the \emph{pMSE} is a function of the predicted probabilities, $\hat p$, which are bounded $\in [0, 1]$, the \emph{pMSE} is bounded $\in [0, 0.25]$. This is true regardless of the range of the data, $X$, so we do not need to assume any kinds of bounds on the data. 

We refer to our method as the \emph{pMSE} Mechanism, since we rely on the \emph{pMSE} for our quality function in the exponential mechanism. Algorithm \ref{algpmseMech} outlines the steps of the \textit{pMSE} mechanism. The main assumption we need is that a reasonable generative model for the data, $g(\theta)$, exists.
\begin{algorithm}
\caption{Sampling DP Synthetic Data via the \emph{pMSE} Mechanism}\label{algpmseMech}
  \begin{algorithmic}[1]
  \Input Original dataset: $X$, chosen value: $\epsilon$, synthesis model: $g(\theta)$, quality function: $u(X, \theta)$
  \item Sample $l$ vectors $\{\tilde\theta_1,...,\tilde\theta_l\}$ from a density proportional to equation \ref{eq:expMech}
  \item For each $\tilde\theta_i$ generate synthetic data set $X^s_i \sim g(\tilde\theta_i)$, giving $m$ total synthetic datasets $\{X^s_1,...,X^s_l\}$ each with the same dimension as $X$
  \item Releasing $\{X^s_1,...,X^s_l\}$ satisfies $(l\epsilon)$-DP
  \end{algorithmic}
\end{algorithm} 

\subsection{Estimating the \textit{pMSE} using Classification and Regression Tree (CART) Models}
A key component to defining our quality function is the classification model used to estimate the predicted probabilities, $\hat p$, used in computing the \emph{pMSE}. We choose the classification trees (\cite{Breiman1984}) fit using the Gini Index, for two primary reasons. First, we need a tight bound on the sensitivity of $u(X, \theta)$. While other machine learning models have been shown to outperform CART in many applications, we would have a far weaker bound on the sensitivity and would need to add much more noise. Secondly, as was shown in \cite{snoke2018utility}, CART models exhibit at least satisfactory performance in determining the distributional similarity. Future work may prove desirable bounds on the sensitivity of the \textit{pMSE} when using stronger classifiers, in which case those models should certainly be adopted.
We use the impurity function known as Gini Index from \cite{Breiman1984}, defined as:
\begin{align}
GI = argmin \,\, \Sigma_{i = 1}^{D + 1}{a_i\Big(1-\frac{a_i}{m_i}\Big)},
\end{align}
where $m_i$ are the total number of observations in each node, $a_i$ are the number of observations labeled $1$ in each node, and $D$ is the total number of nodes. In practice these models are fit in a greedy manner for computational purposes. The process is to make the first optimal split that minimizes the impurity for two nodes, and then to make proceeding splits and adding additional nodes if doing so continues to minimize the impurity according to a chosen cost function. If computation is not a concern, it would also be possible for any fixed $D$ to do a full grid search to determine the optimal $D$ splits that minimize the impurity over $D + 1$ nodes. The difference between globally optimal and greedy trees is important for our theoretical results. In our theoretical results in section \ref{sec:sensBound} we prove the sensitivity bound when trees are fit based on the globally optimal Gini Index, and in our simulations in section \ref{sec:simCompFail} we perform an emperical examination of how frequently the greedy fitting violates our theoretical results.

\section{Theoretical Results for the Sensitivity Bound}
\label{sec:sensBound}
In order to sample from the exponential mechanism, we need to give a bound on the sensitivity of the quality function. The \emph{pMSE} function is naturally bounded, but in Theorem \ref{theorem:sensBound} we prove a much tighter bound.
\begin{theorem}
\label{theorem:sensBound}
\textit{Given $u(X, \theta) = E_\theta[pMSE(X, X^s_\theta) | X, \theta]$ where $pMSE = \Sigma_{i = 1}^{2n} \frac{(\hat p_i - 0.5)^2}{2n}$ with $\hat p_i$ estimated from a classification tree with optimal splits found using the Gini Index. Then
\[
\Delta u = \underset{\theta}{sup}\underset{\delta(X,X^\prime)=1}{sup}|u(X,\theta) - u(X^\prime, \theta)| \leq \frac{1}{n},
\]
where $X,X^s_\theta \in \mathbb{R}^{n \times q}$.}
\end{theorem}

The proof is given in the Appendix. Intuitively, the proof follows from the fact that we can relate the \textit{pMSE} to the Gini Index. We can then bound the change in Gini Index given a change in one row of the input data due to the fact that we are finding the global optimum. We will not suddenly do much better or much worse. In fact, we can quantify exactly how much better or worse we can do, which leads to the bound.

This bound is nice because it decreases with $n$, meaning the noise added decreases as the number of observations increase. 
This bound matches the results derived for the sensitivity of the Gini Index when assuming discrete predictors from \cite{friedman2010data}. Our proof shows this the bound remains the same when using continuous predictors. The result in \cite{friedman2010data} was used for performing classification under differential privacy, rather than producing synthetic data, and we see our extension of the proof to include continuous predictors as a useful side result of this paper.

It is important to note that this proof is for the theoretical case when we can find the optimal partitioning for any number of nodes. 
The greedy method can violate the bound because we can no longer control how much the Gini Index can change after changing one row. While it would be possible to use our method with a full grid search, computationally this is a poor idea. On the other hand, it is necessary in order to satisfy $\epsilon$-DP. An alternative method could be to use adaptive composition, i.e., fit the CART models greedily but in a way that satisfies DP. We could then compose the privacy between fitting the CART model and sampling from the exponential distribution, which we explore in future work. 


\section{Empirical Failure Rate of the Sensitivity Bound}
\label{sec:simCompFail}
These simulations show the empirical rate for which the greedy fitting violates the bound. We can also view the maximum simulated value as an empirical estimate of the sensitivity for this particular dataset, but we are more interested in the failure rate. We generated datasets, $X$, with $q = 2$ and $n = 5000$. $X_1 \sim N(2, 10)$ and $X_2 \sim N(-2.5 + 0.5x_1, 3)$, and we produced $X^\prime$ by taking $X$ and adding random Gaussian noise, $N(0, 25)$, to each variable for one observation. We then drew a synthetic dataset $X^s$ with $X^s_1 \sim N(\theta_1, \theta_2)$ and $X^s_2 \sim N(\theta_3 + \theta_4x^s_1, \theta_5)$ where $\theta_i \sim N(0, 10)$. We estimated the \emph{pMSE} with respect to $X^s$ 
for both $X$ and $X^\prime$ and calculated the difference. Recall the theoretical sensitivity bound is $1/n = 0.0002$, and any values larger than this violate the bound. We repeated this process 1,000,000 times each using CARTs of depths 1, 2, 5, and unlimited for the \emph{pMSE} model. For all trees we included a complexity parameter ($cp$) requiring a certain percentage improvement in order to make an additional split. This parameter is necessary in order to not produce trees that are fully saturated (one terminal node per observation) when there is no depth limitation.
\begin{table}[ht]
\centering
\caption{\label{tab:empirFail} Empirical failure rates of 1,000,000 simulations for the sensitivity bound when using the greedy CART fitting algorithm for different tree sizes and different complexity parameters.}
\begin{tabular}{llc}
  \hline
  Tree Depth & $cp$ & Percentage Violating Bound \\
  \hline
  Depth 1 & $0.01$ & 0.0\% \\ 
  Depth 2 & $0.01$ & 0.3\% \\ 
  Depth 5 & $0.01$ & 0.6\% \\ 
  Depth Unlimited & $0.01$ & 0.7\% \\
  \hline
  Depth 1 & $0.001$ & 0.0\% \\ 
  Depth 2 & $0.001$ & 0.5\% \\ 
  Depth 5 & $0.001$ & 2.0\% \\ 
  Depth Unlimited & $0.001$ & 2.5\% \\ 
  \hline
\end{tabular}
\end{table}

\begin{figure}[ht]
\centering
\includegraphics[width = \textwidth]{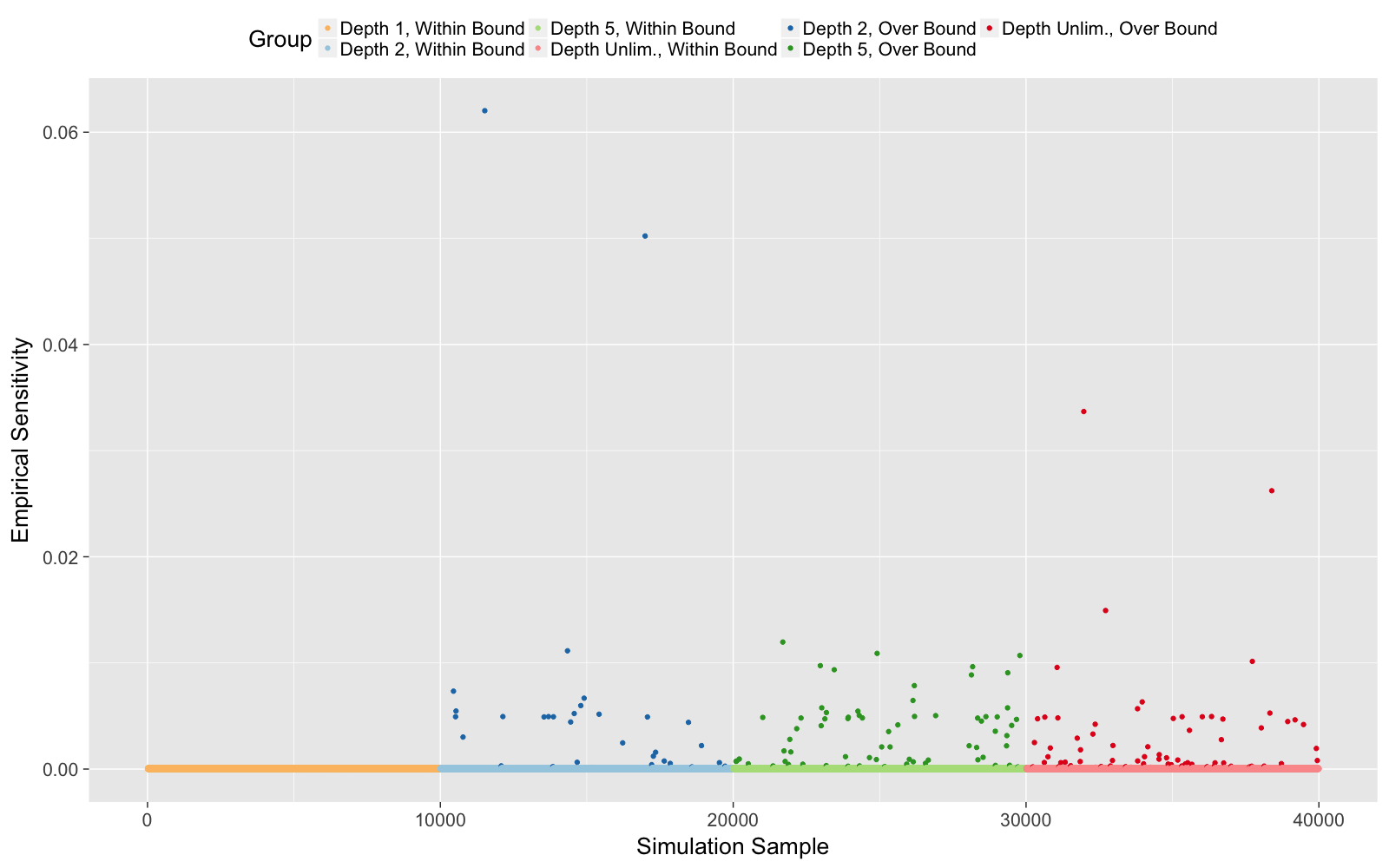}
\caption{\label{fig:sensBound} Random sample of 10,000 simulations for each tree depth. Values shown are differences between $u(X, \theta)$ calculated with X and $X^\prime$. Darker points violate the theoretical bound. $Cp = 1\%$.}
\end{figure}

Figure \ref{fig:sensBound} shows a sample of the simulated empirical sensitivity results. There are four groupings, for the trees with different depths, and darker points denote those violating the theoretical bound due to the greedy fitting algorithm. Table \ref{tab:empirFail} gives the full results. The percentage of simulations which violate the bound increases with tree depth size, and in the unlimited case for $cp = 0.01$ the empirical failure rate is $\leq 1\%$. As expected, there are no results which violate the bound when only one split is made. This confirms our theoretical results because with only one split, greedy is equivalent to optimal, so the bound is never violated in simulation. The empirical sensitivity also depends on the $cp$, so we ran simulations for two different values. A lower $cp$ will lead to larger trees (subject to depth constraints), which means we are making more greedy splits and increasing the chance of violating the bound.

\section{Empirical Evaluation of Differentially Private Linear Regression}
\label{sec:simLM}
In order to assess the practical statistical utility of our method, we ran simulations testing the accuracy of an estimated linear regression model. Our method guarantees maximal distributional similarity of the synthetic data based on the \textit{pMSE} metric, but many researchers may be interested in more specific comparisons such as regression outputs. 

We simulate datasets, $X$, in the same way as in Sections \ref{sec:simCompFail} and \ref{sec:simEmpPmse}. 
Using this data, we regress $X_2$ on $X_1$ and get ordinary least squares (OLS) estimates of the intercept ($\hat\beta_0$) and slope ($\hat\beta_1$) coefficients. We calculate the absolute difference between these estimates and the corresponding estimates we get by fitting the same model with the differentially private methods, i.e., $|\hat\beta - \hat\beta^{priv}|$. Our comparison methods are the noisy Bayesian method from \cite{bowen2016comparative} and the smooth histogram from \cite{wasserman2010statistical}, both described in Section \ref{sec:relatedMethods}. We also compare with methods that do not produce synthetic data but produce DP regression estimates. Specifically, we use the Functional Mechanism of \cite{zhang2012functional} and \cite{awan2018structure}, considering $l_1$ and $l_\infty$ mechanisms, respectively. 
Note that these methods require the data to be bounded in the same way as the noisy Bayes and smooth histogram methods. Finally we compare with estimates from non-DP synthetic data, sampled from the unperturbed BPDD.
\begin{figure}[!ht]
\centering
\includegraphics[width = \textwidth]{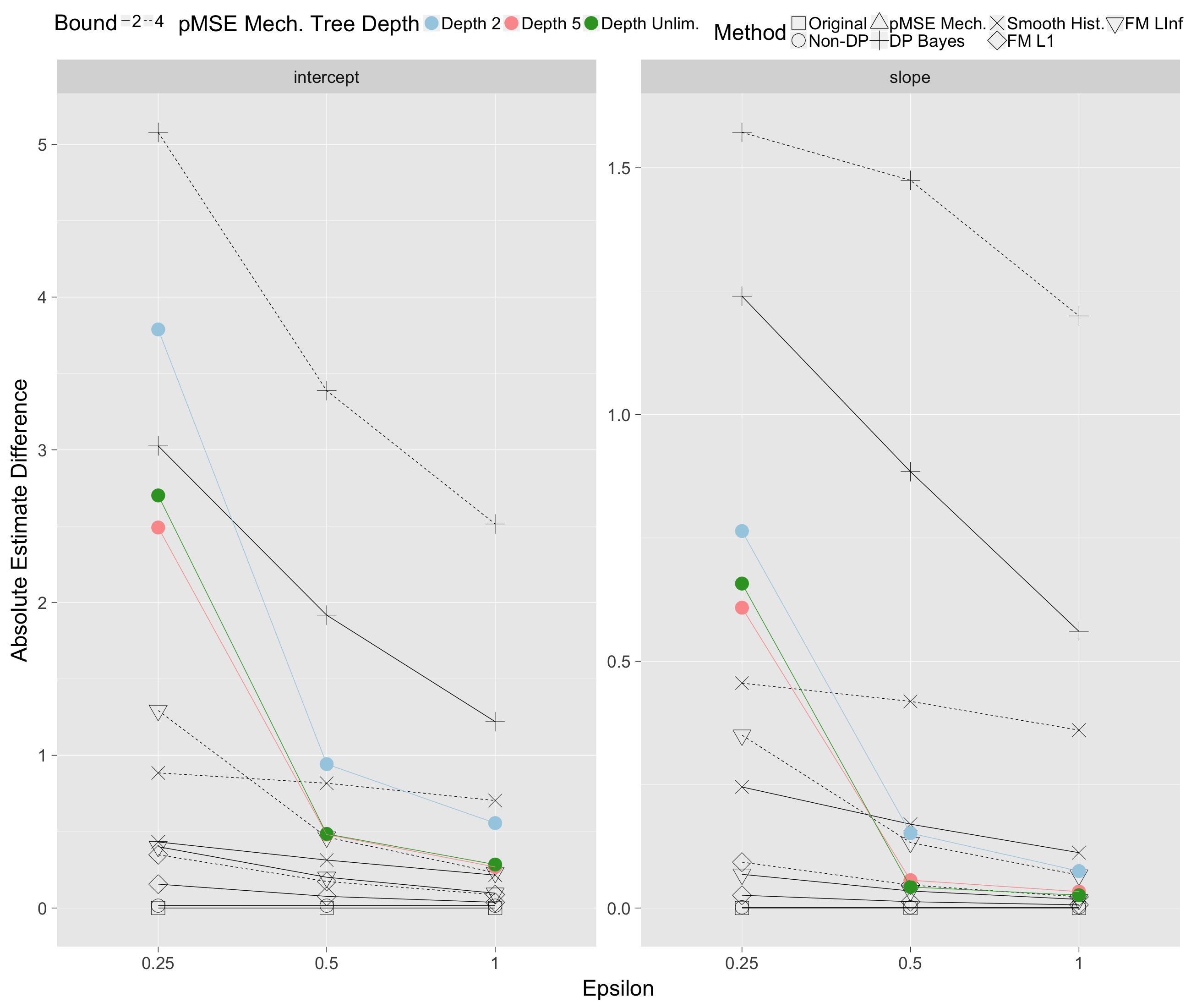}
\caption{\label{fig:2varLine} Lineplots showing the mean simulation results. X-axis indicates different values of $\epsilon$. Lines are also subdivided within methods by the tree depth and the bound.}
\end{figure}

For the \emph{pMSE} mechanism, we carry out the simulations using trees of depths 1, 2, 5, and unlimited with a $cp = 0.01$. We see that the utility significantly improves as we move from depth 1 to 2 and from 2 to 5, but there is little change from 5 to unlimited. This is likely because trees of depth 5 are large enough to evaluate this dataset. The tree size is a potential tuning parameter for future work using this method. Astute readers may have noticed that the unnormalized distribution we propose for the \textit{pMSE} mechanism does not neccesarily exist, since the probability in the tails remains very flat. To fix this, we add a very flat prior, $N(0, 100,000)$, to each of our parameters when sampling. The flatness ensures it does not affect the utility, but by adding it we also ensure the probability in the tails eventually goes to zero.

We run the mechanisms with values $\epsilon \in \{0.25, 0.5, 1\}$. For the \textit{pMSE} mechanism and the noisy BPDD we generate $l = 10$ private datasets each satisfying $(\epsilon / l)$-DP, and for the smooth histogram and functional mechanisms we produce only one output satisfying $\epsilon$-DP. This ensures all mechanisms satisfy the same level of privacy. The non-DP synthetic data method does not guarantee any privacy.

For the noisy Bayes, smooth histogram, and functional mechanism methods, we ran the simulations truncating the data at different assumed bounds. For both variables, we set these bounds at two, four, five, or ten times the standard deviation. Four or five can be thought of as roughly the appropriate bounds, since this is Gaussian data and most observations will fall into those ranges. Two was chosen to be a range that is too narrow and excludes part of the true distribution, and ten was chosen for a looser bound. We see from the results that the smaller bound achieves better results on the regression, even when it is more narrow even than the truth. This is an artefact of the model we chose, and if an even tighter bound was chosen it may have greater adverse effects. The loose bound (10 times) performs quite poorly, since we must add much more noise. Figure \ref{fig:2varLine} visualizes the better performing results (limited only for readability). Full results for all tree sizes and bounds are shown in Figure \ref{fig:2varUtil} in the Appendix.

Overall our method outperforms the other two synthesis methods when using trees of depth 5 or unlimited, regardless of the bounds chosen. Even trees of depth 2 perform roughly the same or better than the other methods.
For deeper trees, our method performs almost as well as the functional mechanism. This is quite encouraging, since that method focuses only on providing regression estimates rather than entire synthetic datasets. For a slight decrease in utility from the functional mechanism, our method provides an entire synthetic dataset, which can be used to fit any number of models using our synthetic data without changing the privacy guarantee. For the functional mechanism, it requires further splitting of the privacy parameter to estimate a different model.

These results show good performance, and further work should consider simulations with larger numbers of variables or a mixture of categorical and continuous variables. We expect our method will only improve against the other methods with more variables, since theoretically our method maximizes similarity on the entire distribution.

\section{Empirical Evaluation of the \textit{pMSE}}
\label{sec:simEmpPmse}
We guarantee theoretical maximization of the \textit{pMSE} for the differentially private synthetic data produced from the \textit{pMSE} mechanism, but as many practitioners know empirical tests often look slightly different from theory. To evaluate this, we ran simulations to estimate the \textit{pMSE} from datasets generated using our method, two other DP synthesis methods, and a standard non-DP synthesis method.
\begin{figure}[ht]
\centering
\includegraphics[width = \textwidth]{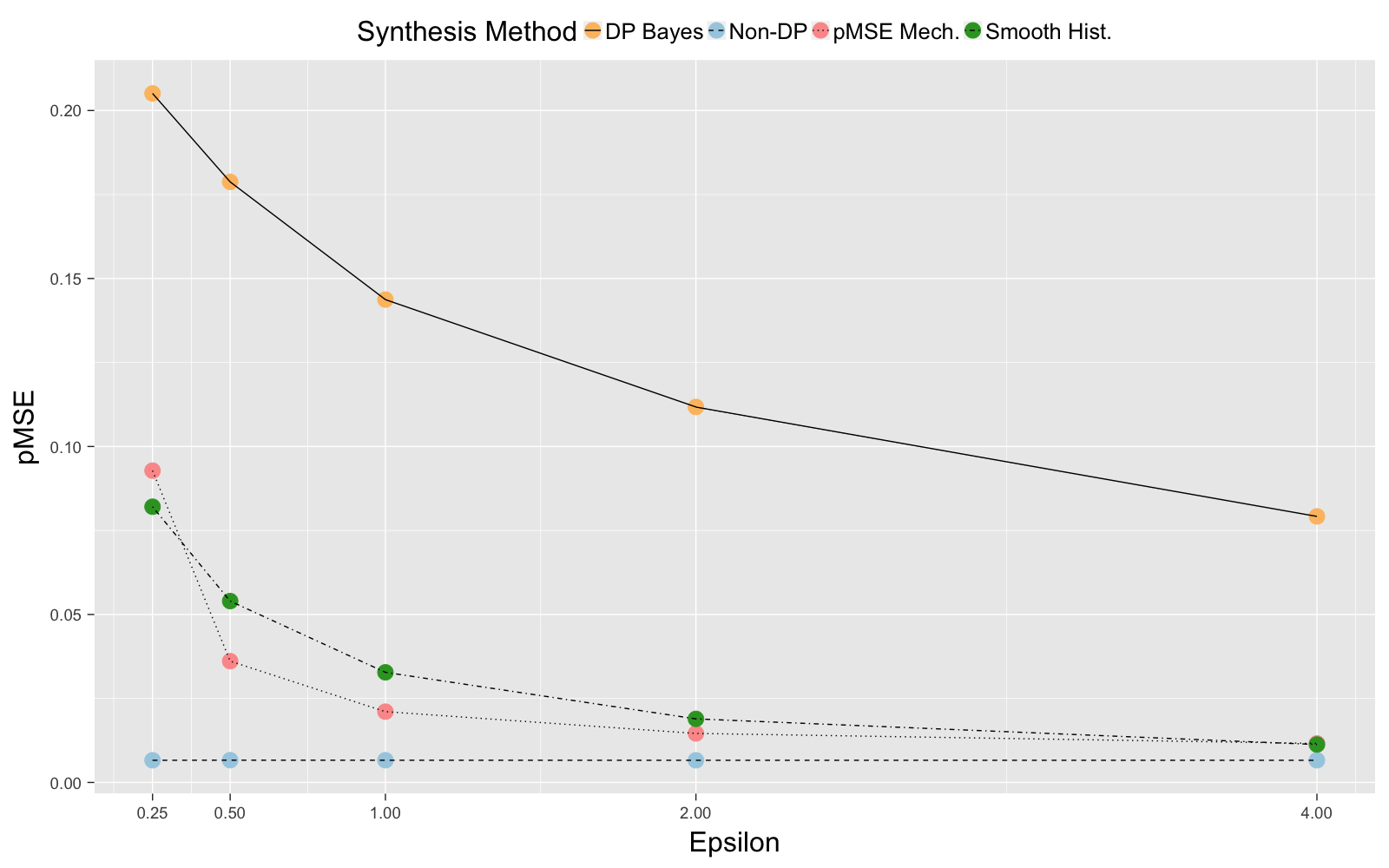}
\caption{\label{fig:empPmse} Simulations results showing the mean \textit{pMSE} calculated using synthetic producing according to four different methods. \textit{pMSE} is calculated from comparison with original data, with values closer to 0 implying higher utility.}
\end{figure}

We again simulate datasets, $X$, in the same way as in Sections \ref{sec:simCompFail} and \ref{sec:simLM}. For each $X$ we generated synthetic datasets $X^s$ and then calculated the \textit{pMSE} using $X$ and $X^s$. Our four synthesis methods were the \textit{pMSE} mechanism, the noisy Bayesian method from \cite{bowen2016comparative}, the smooth histogram from \cite{wasserman2010statistical}, and sampling from the non-differentially private BPPD using fully conditional sequential models. We ran 2,500 simulations each for five different values of $\epsilon \in \{0.25, 0.5, 1, 2, 4\}$. For our method we used CART trees with unlimited depth, and for the other two DP methods we assumed a bound on the data of four times the standard deviation, which is roughly the correct bound given that it is Gaussian data. Figure \ref{fig:empPmse} shows the results for the mean simualtions results and Table \ref{tab:empPmse} shows the full mean and variance of the results.
\setlength{\tabcolsep}{10pt}
\begin{table}[ht]
\centering
\caption{\label{tab:empPmse} Simulation results giving the mean and variance of the \textit{pMSE} values calculated using four different synthesis methods and five different levels of $\epsilon$.}
\begin{tabular}{llcccc}
  \hline
  $\epsilon$ & Simulated Values & Non-DP & \textit{pMSE Mech.} & DP Bayes & Smooth Hist. \\
  \hline
  0.25 & \textit{pMSE} Mean & 0.00660 & 0.09281 & 0.20509 & 0.08206 \\
  & \textit{pMSE} Var. & 8.681e-07 & 1.347e-03 & 3.079e-03 & 2.826e-05 \\
  \hline
  0.5 & \textit{pMSE} Mean & 0.00663 & 0.03610 & 0.17876 & 0.05398 \\
  & \textit{pMSE} Var. & 8.929e-07 & 8.020e-05 & 5.914e-03 & 1.809e-05 \\
  \hline
  1 & \textit{pMSE} Mean & 0.00661 & 0.02107 & 0.14372 & 0.03278 \\
  & \textit{pMSE} Var. & 8.342e-07 & 1.648e-05 & 8.579e-03 & 1.069e-05 \\
  \hline
  2 & \textit{pMSE} Mean & 0.00660 & 0.01459 & 0.11177 & 0.01892 \\
  & \textit{pMSE} Var. & 8.342e-07 & 1.648e-05 & 8.579e-03 & 1.069e-05 \\ 
  \hline
  4 & \textit{pMSE} Mean & 0.00660 & 0.01161 & 0.07919 & 0.01129 \\ 
  & \textit{pMSE} Var. & 8.671e-07 & 3.329e-06 & 9.087e-03 & 5.964e-06 \\ 
  \hline
\end{tabular}
\end{table}

We see that as expected, the \textit{pMSE} mechanism offers either the best or one of the best values of the \textit{pMSE} among the methods guaranteeing differential privacy. Interestingly the smooth histogram method is fairly good as well, even offering comparable values at $\epsilon = 0.25$ or $\epsilon = 4$. The noisy BPPD method on the other hand is quite bad, even at high levels of $\epsilon$, so should be used with  caution. 

We also see that the variance in the estimated \textit{pMSE} values changes quite a bit depending on the method and level of $\epsilon$. Both the \textit{pMSE} mechanism and the smooth histogram show higher variances for either low (0.25) or high (4) values of $\epsilon$, while the noisy BPPD method increases in variance as $\epsilon$ increases. This variance is something to keep in mind both in choosing a protection method and in developing the practical implementation. We could likely improve our current implementation of the \textit{pMSE} mechanism in order to better sample noisy parameters and generate synthetic data with less variance in the resulting \textit{pMSE}. On the other hand, it should also be expected that the variance decreases some as $\epsilon$ grows because we are adding less noise through the privacy mechanism.

Comparing the DP methods to the traditional synthetic data approach, we see that the best method at $\epsilon = 4$ produces an average \textit{pMSE} roughly two times that from the non-DP synthesis method, which is producing synthetic data from the correct generative model. This is actually quite good considering we are adding the strong guarantee of DP. Even for $\epsilon = 1$ our method produces \textit{pMSE} values only roughly three times that of the non-DP synthetic data.

\section{Conclusions and Future Work}
The \emph{pMSE} mechanism we propose provides a novel flexible method to produce high-quality synthetic datasets guaranteeing $\epsilon$-DP. By sampling generative model parameters from the exponential mechanism and using the \emph{pMSE} as our quality function, we produce synthetic data with maximal distributional similarity to the original data. By using the \emph{pMSE}, we ensure the sensitivity depends neither on the dimension nor the range of the data, and the bound decreases as we increase the sample size. This allows us to use this mechanism for continuous data, and the amount of noise we add will not grow with the dimension (apart from sampling from a more complex distribution).

Our simulations in Sections \ref{sec:simLM} and \ref{sec:simEmpPmse} confirm that the \textit{pMSE} mechanism generally performs as well or better than the other standard DP synthesis methods. In the case of linear regression the \textit{pMSE} mechanism even performs roughly as well as methods that produce estimates of regression coefficients only rather than entire synthetic datasets. In the case of the empirical \textit{pMSE}, as expected our method performs worse than non-DP synthetic data, but the utility cost seems reasonable for the privacy gain. 

The \emph{pMSE} mechanism relies on defining an appropriate form for the generative distribution from which to draw synthetic values. It is possible to misspecific this model, which would lead to poor utility. This is one drawback of the synthetic approach as opposed to simply adding noise. Fortnately, this aspect has been addressed in great detail in the synthetic data literature, so we feel that finding an appropriate model is possible without too much difficulty. 

Our primary limitation is the computational feasibility to ensure the theoretical sensitivity bound. From the empirical simulations we saw that the bound does not always hold when using typical greedy fitting algorithms. Fitting the models using the global optimum would ensure the theoretical bound and guarantee $\epsilon$-DP. Proposals have been made to carry out machine learning using global optimums, such as \cite{bertsimas2017optimal}, so methods may exist to aid the computation.

An alternative implementation of our method would be to consider fitting the CART models in a way that satisfies $\epsilon$-DP and then composing this $\epsilon$ with that from sampling from the \textit{pMSE} mechanism. This is similar to the approach of \cite{li2018highdimHist}.
 This is desirable because we could use any standard CART software to implement the method.
 Other future work could consider using different impurity measures than the Gini Index, deriving measures of choosing the best tree size, or best practices for sampling from the unnormalized density we get through the exponential mechanism.

\bibliographystyle{chicago}
\bibliography{dissertation}

\begin{thebibliography}{}

\bibitem[\protect\citeauthoryear{Abowd and Vilhuber}{Abowd and
  Vilhuber}{2008}]{abowd2008protective}
Abowd, J.~M. and L.~Vilhuber (2008).
\newblock How protective are synthetic data?
\newblock In {\em International Conference on Privacy in Statistical
  Databases}, pp.\  239--246. Springer.

\bibitem[\protect\citeauthoryear{Awan and Slavkovic}{Awan and
  Slavkovic}{2018}]{awan2018structure}
Awan, J. and A.~Slavkovic (2018).
\newblock Structure and sensitivity in differential privacy: Comparing k-norm
  mechanisms.
\newblock {\em arXiv preprint arXiv:1801.09236\/}.

\bibitem[\protect\citeauthoryear{Barak, Chaudhuri, Dwork, Kale, McSherry, and
  Talwar}{Barak et~al.}{2007}]{barak2007privacy}
Barak, B., K.~Chaudhuri, C.~Dwork, S.~Kale, F.~McSherry, and K.~Talwar (2007).
\newblock Privacy, accuracy, and consistency too: a holistic solution to
  contingency table release.
\newblock In {\em Proceedings of the twenty-sixth ACM SIGMOD-SIGACT-SIGART
  symposium on Principles of database systems}, pp.\  273--282. ACM.

\bibitem[\protect\citeauthoryear{Bertsimas and Dunn}{Bertsimas and
  Dunn}{2017}]{bertsimas2017optimal}
Bertsimas, D. and J.~Dunn (2017).
\newblock Optimal classification trees.
\newblock {\em Machine Learning\/}~{\em 106\/}(7), 1039--1082.

\bibitem[\protect\citeauthoryear{Blum, Dwork, McSherry, and Nissim}{Blum
  et~al.}{2005}]{blum2005practical}
Blum, A., C.~Dwork, F.~McSherry, and K.~Nissim (2005).
\newblock Practical privacy: the sulq framework.
\newblock In {\em Proceedings of the twenty-fourth ACM SIGMOD-SIGACT-SIGART
  symposium on Principles of database systems}, pp.\  128--138. ACM.

\bibitem[\protect\citeauthoryear{Bowen and Liu}{Bowen and
  Liu}{2016}]{bowen2016comparative}
Bowen, C.~M. and F.~Liu (2016).
\newblock Comparative study of differentially private data synthesis methods.
\newblock {\em arXiv preprint arXiv:1602.01063\/}.

\bibitem[\protect\citeauthoryear{Breiman, Friedman, Olshen, and Stone}{Breiman
  et~al.}{1984}]{Breiman1984}
Breiman, L., J.~H. Friedman, R.~A. Olshen, and C.~J. Stone (1984).
\newblock {\em Classification and Regression Trees}.
\newblock Belmont, Wadsworth, CA.

\bibitem[\protect\citeauthoryear{Bun and Steinke}{Bun and
  Steinke}{2016}]{bun2016concentrated}
Bun, M. and T.~Steinke (2016).
\newblock Concentrated differential privacy: Simplifications, extensions, and
  lower bounds.
\newblock In {\em Theory of Cryptography Conference}, pp.\  635--658. Springer.

\bibitem[\protect\citeauthoryear{Charest}{Charest}{2011}]{charest2011can}
Charest, A.-S. (2011).
\newblock How can we analyze differentially-private synthetic datasets?
\newblock {\em Journal of Privacy and Confidentiality\/}~{\em 2\/}(2), 3.

\bibitem[\protect\citeauthoryear{Chaudhuri, Sarwate, and Sinha}{Chaudhuri
  et~al.}{2012}]{chaudhuri2012near}
Chaudhuri, K., A.~Sarwate, and K.~Sinha (2012).
\newblock Near-optimal differentially private principal components.
\newblock In {\em Advances in Neural Information Processing Systems}, pp.\
  989--997.

\bibitem[\protect\citeauthoryear{Drechsler}{Drechsler}{2011}]{drechsler2011synthetic}
Drechsler, J. (2011).
\newblock {\em Synthetic datasets for statistical disclosure control: theory
  and implementation}, Volume 201.
\newblock Springer Science \& Business Media.

\bibitem[\protect\citeauthoryear{Dwork, Kenthapadi, McSherry, Mironov, and
  Naor}{Dwork et~al.}{2006}]{dwork2006our}
Dwork, C., K.~Kenthapadi, F.~McSherry, I.~Mironov, and M.~Naor (2006).
\newblock Our data, ourselves: Privacy via distributed noise generation.
\newblock In {\em Annual International Conference on the Theory and
  Applications of Cryptographic Techniques}, pp.\  486--503. Springer.

\bibitem[\protect\citeauthoryear{Dwork, McSherry, Nissim, and Smith}{Dwork
  et~al.}{2006}]{dwork2006calibrating}
Dwork, C., F.~McSherry, K.~Nissim, and A.~Smith (2006).
\newblock Calibrating noise to sensitivity in private data analysis.
\newblock In {\em Theory of cryptography}, pp.\  265--284. Springer.

\bibitem[\protect\citeauthoryear{Dwork, Naor, Pitassi, Rothblum, and
  Yekhanin}{Dwork et~al.}{2010}]{dwork2010pan}
Dwork, C., M.~Naor, T.~Pitassi, G.~N. Rothblum, and S.~Yekhanin (2010).
\newblock Pan-private streaming algorithms.
\newblock In {\em ICS}, pp.\  66--80.

\bibitem[\protect\citeauthoryear{Dwork, Roth, et~al.}{Dwork
  et~al.}{2014}]{dwork2014algorithmic}
Dwork, C., A.~Roth, et~al. (2014).
\newblock The algorithmic foundations of differential privacy.
\newblock {\em Foundations and Trends{\textregistered} in Theoretical Computer
  Science\/}~{\em 9\/}(3--4), 211--407.

\bibitem[\protect\citeauthoryear{Dwork and Rothblum}{Dwork and
  Rothblum}{2016}]{dwork2016concentrated}
Dwork, C. and G.~N. Rothblum (2016).
\newblock Concentrated differential privacy.
\newblock {\em arXiv preprint arXiv:1603.01887\/}.

\bibitem[\protect\citeauthoryear{Friedman and Schuster}{Friedman and
  Schuster}{2010}]{friedman2010data}
Friedman, A. and A.~Schuster (2010).
\newblock Data mining with differential privacy.
\newblock In {\em Proceedings of the 16th ACM SIGKDD international conference
  on Knowledge discovery and data mining}, pp.\  493--502. ACM.

\bibitem[\protect\citeauthoryear{Kapralov and Talwar}{Kapralov and
  Talwar}{2013}]{kapralov2013differentially}
Kapralov, M. and K.~Talwar (2013).
\newblock On differentially private low rank approximation.
\newblock In {\em Proceedings of the twenty-fourth annual ACM-SIAM symposium on
  Discrete algorithms}, pp.\  1395--1414. SIAM.

\bibitem[\protect\citeauthoryear{Karwa, Krivitsky, and Slavkovi{\'c}}{Karwa
  et~al.}{2017}]{karwa2017sharing}
Karwa, V., P.~N. Krivitsky, and A.~B. Slavkovi{\'c} (2017).
\newblock Sharing social network data: differentially private estimation of
  exponential family random-graph models.
\newblock {\em Journal of the Royal Statistical Society: Series C (Applied
  Statistics)\/}~{\em 66\/}(3), 481--500.

\bibitem[\protect\citeauthoryear{Karwa, Slavkovi{\'c}, et~al.}{Karwa
  et~al.}{2016}]{karwa2016inference}
Karwa, V., A.~Slavkovi{\'c}, et~al. (2016).
\newblock Inference using noisy degrees: Differentially private $\beta $-model
  and synthetic graphs.
\newblock {\em The Annals of Statistics\/}~{\em 44\/}(1), 87--112.

\bibitem[\protect\citeauthoryear{Kasiviswanathan, Lee, Nissim, Raskhodnikova,
  and Smith}{Kasiviswanathan et~al.}{2011}]{kasiviswanathan2011can}
Kasiviswanathan, S.~P., H.~K. Lee, K.~Nissim, S.~Raskhodnikova, and A.~Smith
  (2011).
\newblock What can we learn privately?
\newblock {\em SIAM Journal on Computing\/}~{\em 40\/}(3), 793--826.

\bibitem[\protect\citeauthoryear{Kifer, Smith, and Thakurta}{Kifer
  et~al.}{2012}]{kifer2012private}
Kifer, D., A.~Smith, and A.~Thakurta (2012).
\newblock Private convex empirical risk minimization and high-dimensional
  regression.
\newblock In {\em Conference on Learning Theory}, pp.\  25--1.

\bibitem[\protect\citeauthoryear{Li, Karwa, Slavkovi\'{c}, and Steorts}{Li
  et~al.}{2018}]{li2018highdimHist}
Li, Karwa, Slavkovi\'{c}, and Steorts (2018).
\newblock Release of differentially private high dimensional histograms.
\newblock {\em Pre-print\/}.

\bibitem[\protect\citeauthoryear{Li, Hay, Rastogi, Miklau, and McGregor}{Li
  et~al.}{2010}]{li2010optimizing}
Li, C., M.~Hay, V.~Rastogi, G.~Miklau, and A.~McGregor (2010).
\newblock Optimizing linear counting queries under differential privacy.
\newblock In {\em Proceedings of the twenty-ninth ACM SIGMOD-SIGACT-SIGART
  symposium on Principles of database systems}, pp.\  123--134. ACM.

\bibitem[\protect\citeauthoryear{Machanavajjhala, Kifer, Abowd, Gehrke, and
  Vilhuber}{Machanavajjhala et~al.}{2008}]{machanavajjhala2008privacy}
Machanavajjhala, A., D.~Kifer, J.~Abowd, J.~Gehrke, and L.~Vilhuber (2008).
\newblock Privacy: Theory meets practice on the map.
\newblock In {\em Data Engineering, 2008. ICDE 2008. IEEE 24th International
  Conference on}, pp.\  277--286. IEEE.

\bibitem[\protect\citeauthoryear{McSherry and Talwar}{McSherry and
  Talwar}{2007}]{mcsherry2007mechanism}
McSherry, F. and K.~Talwar (2007).
\newblock Mechanism design via differential privacy.
\newblock In {\em Foundations of Computer Science, 2007. FOCS'07. 48th Annual
  IEEE Symposium on}, pp.\  94--103. IEEE.

\bibitem[\protect\citeauthoryear{McSherry}{McSherry}{2009}]{mcsherry2009privacy}
McSherry, F.~D. (2009).
\newblock Privacy integrated queries: an extensible platform for
  privacy-preserving data analysis.
\newblock In {\em Proceedings of the 2009 ACM SIGMOD International Conference
  on Management of data}, pp.\  19--30. ACM.

\bibitem[\protect\citeauthoryear{Nissim, Raskhodnikova, and Smith}{Nissim
  et~al.}{2007}]{nissim2007smooth}
Nissim, K., S.~Raskhodnikova, and A.~Smith (2007).
\newblock Smooth sensitivity and sampling in private data analysis.
\newblock In {\em Proceedings of the thirty-ninth annual ACM symposium on
  Theory of computing}, pp.\  75--84. ACM.

\bibitem[\protect\citeauthoryear{Nissim, Steinke, Wood, Altman, Bembenek, Bun,
  Gaboardi, O’Brien, and Vadhan}{Nissim
  et~al.}{2017}]{nissim2017differential}
Nissim, K., T.~Steinke, A.~Wood, M.~Altman, A.~Bembenek, M.~Bun, M.~Gaboardi,
  D.~R. O’Brien, and S.~Vadhan (2017).
\newblock Differential privacy: A primer for a non-technical audience.

\bibitem[\protect\citeauthoryear{Raab, Nowok, and Dibben}{Raab
  et~al.}{2017}]{raab2017practical}
Raab, G.~M., B.~Nowok, and C.~Dibben (2017).
\newblock Practical data synthesis for large samples.
\newblock {\em Journal of Privacy and Confidentiality\/}~{\em 7\/}(3), 4.

\bibitem[\protect\citeauthoryear{Raghunathan, Reiter, and Rubin}{Raghunathan
  et~al.}{2003}]{RRR_2003}
Raghunathan, T.~E., J.~P. Reiter, and D.~B. Rubin (2003).
\newblock Multiple imputation for statistical disclosure limitation.
\newblock {\em Journal of Official Statistics\/}~{\em 19\/}(1), 1--17.

\bibitem[\protect\citeauthoryear{Reiter}{Reiter}{2002}]{Reiter_2010}
Reiter, J.~P. (2002).
\newblock Satisfying disclosure restrictions with synthetic data sets.
\newblock {\em Journal of Official Statistics\/}~{\em 18\/}(4), 531--544.

\bibitem[\protect\citeauthoryear{Reiter}{Reiter}{2005}]{reiter_cart}
Reiter, J.~P. (2005).
\newblock Using cart to generate partially synthetic, public use microdata.
\newblock {\em Journal of Official Statistics\/}~{\em 21\/}(3), 441--462.

\bibitem[\protect\citeauthoryear{Snoke, Raab, Nowok, Dibben, and
  Slavkovi\'{c}}{Snoke et~al.}{2018}]{snoke2018utility}
Snoke, J., G.~M. Raab, B.~Nowok, C.~Dibben, and A.~Slavkovi\'{c} (2018).
\newblock General and specific utility for synthetic data.
\newblock {\em Journal of the Royal Statistical Society Series A: Statistics in
  Society\/}.

\bibitem[\protect\citeauthoryear{Wang, Lei, and Fienberg}{Wang
  et~al.}{2016}]{wang2016average}
Wang, Y.-X., J.~Lei, and S.~E. Fienberg (2016).
\newblock On-average kl-privacy and its equivalence to generalization for
  max-entropy mechanisms.
\newblock In {\em International Conference on Privacy in Statistical
  Databases}, pp.\  121--134. Springer.

\bibitem[\protect\citeauthoryear{Wasserman and Zhou}{Wasserman and
  Zhou}{2010}]{wasserman2010statistical}
Wasserman, L. and S.~Zhou (2010).
\newblock A statistical framework for differential privacy.
\newblock {\em Journal of the American Statistical Association\/}~{\em
  105\/}(489), 375--389.

\bibitem[\protect\citeauthoryear{Woo, Reiter, Oganian, and Karr}{Woo
  et~al.}{2009}]{WooProp}
Woo, M.-J., J.~P. Reiter, A.~Oganian, and A.~F. Karr (2009).
\newblock Global measures of data utility for microdata masked for disclosure
  limitation.
\newblock {\em Journal of Privacy and Confidentiality\/}~{\em 1}, 111--124.

\bibitem[\protect\citeauthoryear{Xu, Zhang, Xiao, Yang, Yu, and Winslett}{Xu
  et~al.}{2013}]{xu2013differentially}
Xu, J., Z.~Zhang, X.~Xiao, Y.~Yang, G.~Yu, and M.~Winslett (2013).
\newblock Differentially private histogram publication.
\newblock {\em The VLDB Journal\/}~{\em 22\/}(6), 797--822.

\bibitem[\protect\citeauthoryear{Zhang, Cormode, Procopiuc, Srivastava, and
  Xiao}{Zhang et~al.}{2017}]{zhang2017privbayes}
Zhang, J., G.~Cormode, C.~M. Procopiuc, D.~Srivastava, and X.~Xiao (2017).
\newblock Privbayes: Private data release via bayesian networks.
\newblock {\em ACM Transactions on Database Systems (TODS)\/}~{\em 42\/}(4),
  25.

\bibitem[\protect\citeauthoryear{Zhang, Zhang, Xiao, Yang, and Winslett}{Zhang
  et~al.}{2012}]{zhang2012functional}
Zhang, J., Z.~Zhang, X.~Xiao, Y.~Yang, and M.~Winslett (2012).
\newblock Functional mechanism: regression analysis under differential privacy.
\newblock {\em Proceedings of the VLDB Endowment\/}~{\em 5\/}(11), 1364--1375.

\end{thebibliography}

\section{Appendix: Proof of Theorem 4.1}
\label{sec:appendixProof}
\begin{proof}
We first show that using the expected value, and approximating it, can be bounded above by the supremum across all possible datasets $X^s$ generated using $\theta$. 
\begin{align}
\begin{split}
\Delta u = \underset{\theta}{sup} \; \underset{\delta(X,X^\prime)=1}{sup} \; |u(X, \theta) - u(X^\prime, \theta)|
\end{split}
\end{align}
can be rewritten as
\begin{align}
\begin{split}
\Delta u = \underset{\theta}{sup} \; \underset{\delta(X,X^\prime)=1}{sup} \; |E_\theta[pMSE(X, X^s_\theta)|X,\theta] - E_\theta[pMSE(X^\prime, X^s_\theta)|X,\theta]|
\end{split}
\end{align}
where $u(X, \theta) = E_\theta[pMSE(X, X^s_\theta)|X, \theta]$. Since the absolute value is a convex function, we can apply Jensen's inequality and get
\begin{align}
\begin{split}
\leq \underset{\theta}{sup} \; \underset{\delta(X,X^\prime)=1}{sup} \; E_\theta[|pMSE(X, X^s_\theta) - pMSE(X^\prime, X^s_\theta)||X,\theta].
\end{split}
\end{align}
Then by taking the supremum over any data set $X^s_\theta$, we obtain
\begin{align}
\begin{split}
\leq \underset{X^s_\theta}{sup} \; \underset{\delta(X,X^\prime)=1}{sup} \; |pMSE(X,X^s_\theta) - pMSE(X^\prime, X^s_\theta)|.
\end{split}
\end{align}
This also bounds our approximation of the expected value that we propose to use in practice, since the supremum is also greater than or equal to the sample mean. 

Now writing this explicitly in terms of the CART model, we get
\begin{align}
\begin{split}
\underset{a_i, \; m_i, \; a_i^\prime, \; m_i^\prime}{sup} \; \frac{1}{2n}\Bigg|\Sigma_{i = 1}^{D + 1} m_i\Big(\frac{a_i}{m_i} - 0.5\Big)^2 - m_i^\prime\Big(\frac{a_i^\prime}{m_i^\prime} - 0.5\Big)^2\Bigg|
\end{split}
\end{align}
where $a_i$, $m_i$, and $D$ are defined as before, and $a_i^\prime$ and $m_i^\prime$ are the corresponding values for the model fit using $X^\prime$. Expanding this we get
\begin{align}
\begin{split}
\underset{a_i, \; m_i, \; a_i^\prime, \; m_i^\prime}{sup} \; \frac{1}{2n}\Bigg|\Sigma_{i = 1}^{D + 1} \Big(\frac{a_i^2}{m_i} - a_i - 0.25m_i\Big) - \Big(\frac{a_i^{\prime 2}}{m_i^\prime} - a_i^\prime - 0.25m_i^\prime\Big)\Bigg|
\end{split}
\end{align}
and we can cancel the third terms because $\Sigma_{i=1}^{D+1}m_i = \Sigma_{i=1}^{D+1}m_i^\prime$. When we multiple by $2n$, the remaining inside term is equivalent to
the sensitivity of the impurity, i.e.,
\begin{align}
\begin{split}
\underset{a_i, \; m_i, \; a_i^\prime, \; m_i^\prime}{sup} \; \Bigg|GI(X, X^s, D) - GI(X^\prime, X^s, D) \Bigg| = \Delta GI
\end{split}
\end{align}
By bounding the impurity, we bound the \emph{pMSE}. We can rewrite the above as
\begin{align}
\begin{split}
\Bigg|\underset{D}{min} \; GI(X, X^s, D) - \underset{D}{min} \; GI(X^\prime, X^s, D) \Bigg|
\end{split}
\end{align}
since the optimal CART model finds the minimum impurity across any $D$. The greatest possible difference then is the difference between these two optimums. And we can bound this above by
\begin{align}
\begin{split}
\leq \underset{D}{sup} \; \Bigg|GI(X, X^s, D) - GI(X^\prime, X^s, D) \Bigg|.
\end{split}
\end{align}
Let $X^{comb}$ and $X^{\prime comb}$ be the combined data matrices as described in Algorithm \ref{algDPpmse}, including the $0, 1$ outcome variable. Recall that only one record has changed between $X^{comb}$ and $X^{\prime comb}$ (total number of records staying fixed), and it is labeled $0$. We know that for a given $D$ optimal split points producing $D + 1$ nodes on $X^{comb}$, there are $a_i$ records labeled $1$ and $\tilde m_i$ total records in each bin, such that $\exists \; j \ne k \ne l_1 \ne ... \ne l_{D-1} \; s.t. \; \tilde m_j - m_j = m_k - \tilde m_k = 1, \; \tilde m_{l_v} = m_{l_v}$ for $v = \{1,..., D - 1\}$. In the same way, for a given $D$ optimal split points producing $D + 1$ nodes on $X^{\prime comb}$, there are $a^\prime_i$ records labeled $1$ and $\tilde m^\prime_i$ total records in each bin, such that $\exists \; j^\prime \ne k^\prime \ne l^\prime_1 \ne ... \ne l^\prime_{D-1} \; s.t. \; \tilde m^\prime_{j^\prime} - m^\prime_{j^\prime} = m^\prime_{k^\prime} - \tilde m^\prime_{k^\prime} = 1, \; \tilde m^\prime_{l^\prime_v} = m^\prime_{l^\prime_v}$ for $v = \{1,..., D - 1\}$. What this simply means is that after changing one record, the discrete counts in the nodes change by at most one in two of the nodes and does not change in the other $D - 1$ nodes. 

Due to the fact that the CART model produces the $D$ splits that minimize the impurity, we know both that
\begin{align}
\begin{split}\label{eq:impIneq1}
\Sigma_{i = 1}^{D + 1}{a^\prime_i\Big(1-\frac{a^\prime_i}{m^\prime_i}\Big)} \leq \Sigma_{i = 1}^{D + 1}{a_i\Big(1-\frac{a_i}{\tilde m_i}\Big)}
\end{split}
\end{align}
and
\begin{align}
\begin{split}\label{eq:impIneq2}
\Sigma_{i = 1}^{D + 1}{a_i\Big(1-\frac{a_i}{m_i}\Big)} \leq \Sigma_{i = 1}^{D + 1}{a^\prime_i\Big(1-\frac{a^\prime_i}{\tilde m^\prime_i}\Big)}.
\end{split}
\end{align}
The inequality \ref{eq:impIneq1} implies that after changing one record, if new split points are chosen, the impurity must be equivalent or better than simply keeping the previous splits and changing the counts. The inequality \ref{eq:impIneq2} implies that the first split points chosen must be equivalent or better than using the new splits with the changed counts. If this were not the case, the first split points would have never been made in the first place. These lead to the final step. 

Because we have an absolute value, we consider two cases.

\underline{Case 1: $GI(X, X^s, D) \geq GI(X^\prime, X^s, D)$}
\begin{multline}
\underset{D}{sup} \; \Bigg|\Sigma_{i = 1}^{D + 1}{a_i\Big(1-\frac{a_i}{m_i}\Big)} - \Sigma_{i = 1}^{D + 1}{a_i^\prime\Big(1-\frac{a^\prime_i}{m^\prime_i}\Big)} \Bigg| \leq \\
\underset{D}{sup} \; \Bigg|\Sigma_{i = 1}^{D + 1}{a^\prime_i\Big(1-\frac{a^\prime_i}{\tilde m^\prime_i}\Big)} - \Sigma_{i = 1}^{D + 1}{a_i^\prime\Big(1-\frac{a^\prime_i}{m^\prime_i}\Big)}\Bigg| = \\
\Bigg|a^\prime_{j^\prime}\Big(1 - \frac{a^\prime_{j^\prime}}{\tilde m^\prime_{j^\prime}}\Big) - a^\prime_{j^\prime}\Big(1 - \frac{a^\prime_{j^\prime}}{m^\prime_{j^\prime}}\Big) + a^\prime_{k^\prime}\Big(1 - \frac{a^\prime_{k^\prime}}{\tilde m^\prime_{k^\prime}}\Big) - a^\prime_{k^\prime}\Big(1 - \frac{a^\prime_{k^\prime}}{m^\prime_{k^\prime}}\Big)\Bigg| = \\
\Bigg|\frac{a^{\prime 2}_{j^\prime}(\tilde m^\prime_{j^\prime} - m^\prime_{j^\prime})}{\tilde m^\prime_{j^\prime} m^\prime_{j^\prime}} + \frac{a^{\prime 2}_{k^\prime}(\tilde m^\prime_{k^\prime} - m^\prime_{k^\prime})}{\tilde m^\prime_{k^\prime} m^\prime_{k^\prime}}\Bigg|  =
\Bigg|\frac{a^{\prime 2}_{j^\prime}}{\tilde m^\prime_{j^\prime} m^\prime_{j^\prime}} - \frac{a^{\prime 2}_{k^\prime}}{\tilde m^\prime_{k^\prime} m^\prime_{k^\prime}}\Bigg|
\leq 2
\end{multline}
The last step we know because $a_i \leq m_i$, and $\frac{n^2}{n(n-1)} \leq 2$.

\underline{Case 2: $GI(X^\prime, X^s, D) \geq GI(X, X^s, D)$}
\begin{multline}
\underset{D}{sup} \; \Bigg|\Sigma_{i = 1}^{D + 1}{a_i\Big(1-\frac{a_i}{m_i}\Big)} - \Sigma_{i = 1}^{D + 1}{a_i^\prime\Big(1-\frac{a^\prime_i}{m^\prime_i}\Big)} \Bigg| \leq \\
\underset{D}{sup} \; \Bigg|\Sigma_{i = 1}^{D + 1}{a_i\Big(1-\frac{a_i}{\tilde m_i}\Big)} - \Sigma_{i = 1}^{D + 1}{a_i\Big(1-\frac{a_i}{m_i}\Big)}\Bigg| = \\
\Bigg|a_{j}\Big(1 - \frac{a_{j}}{\tilde m_{j}}\Big) - a_{j}\Big(1 - \frac{a_{j}}{m_{j}}\Big) + a_{k}\Big(1 - \frac{a_{k}}{\tilde m_{k}}\Big) - a_{k}\Big(1 - \frac{a_{k}}{m_{k}}\Big)\Bigg| = \\
\Bigg|\frac{a^{\prime 2}_{j}(\tilde m_{j} - m_{j})}{\tilde m_{j} m_{j}} + \frac{a^{\prime 2}_{k}(\tilde m_{k} - m_{k})}{\tilde m_{k} m_{k}}\Bigg|  =
\Bigg|\frac{a^{\prime 2}_{j}}{\tilde m_{j} m_{j}} - \frac{a^{\prime 2}_{k}}{\tilde m_{k} m_{k}}\Bigg|
\leq 2
\end{multline}

Finally, this gives us $\Delta GI \leq 2 \implies \frac{\Delta GI}{2n} = \Delta u \leq \frac{1}{n}$.
\end{proof}

\section{Appendix: Full Simulation Results}
\label{sec:appendixSims}
\begin{sidewaysfigure}[!ht]
\centering
\includegraphics[scale = 0.175]{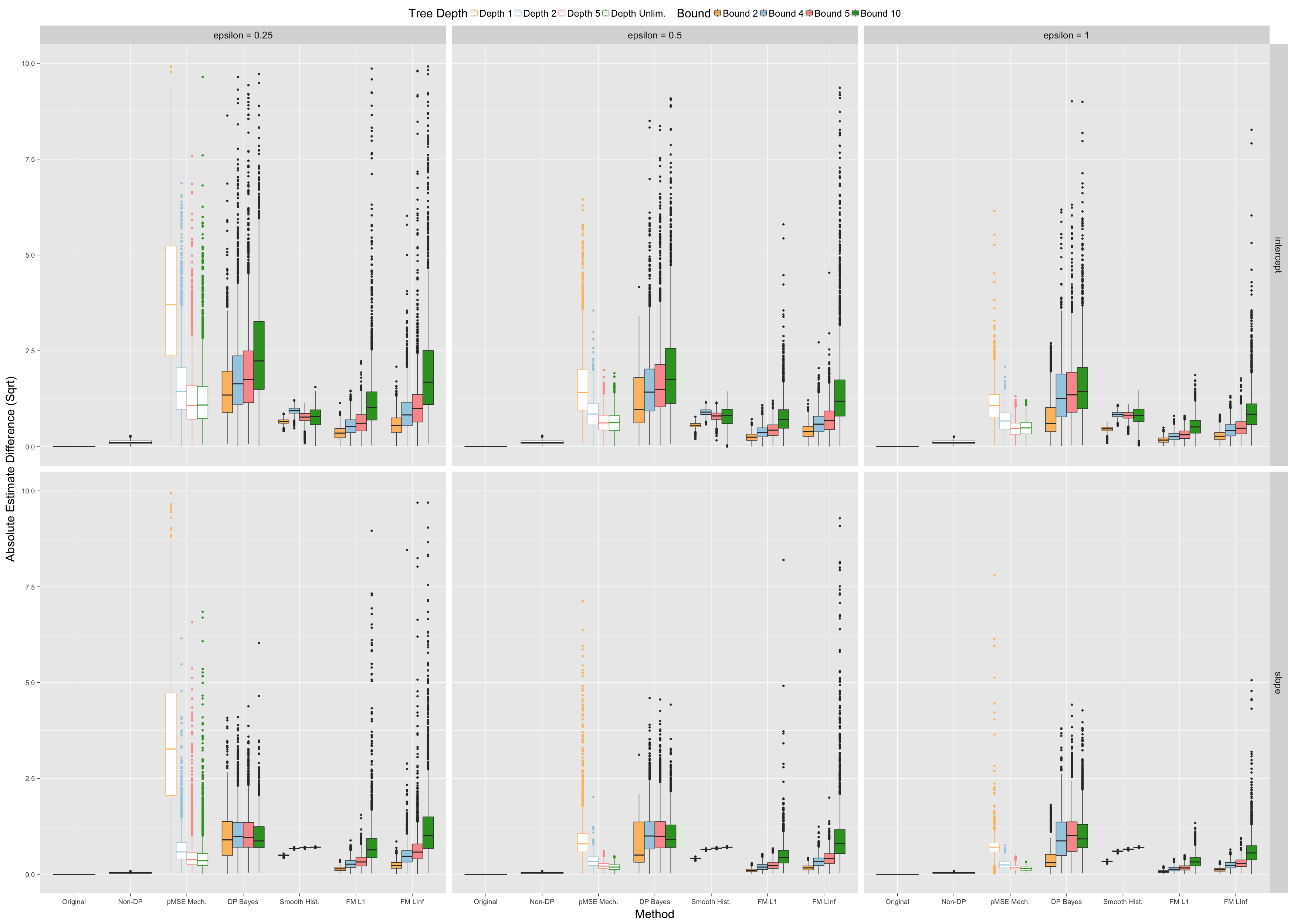}
\caption{\label{fig:2varUtil} Boxplots showing simulation results. The rows indicate the different coefficients, and the columns indicate different values of $\epsilon$. Boxplots are also subdivided within methods by the tree depth (for the \emph{pMSE} mechanism method) and the bound (for others).}
\end{sidewaysfigure}

\end{document}